%% file: main.tex
\documentclass[runningheads]{llncs}
\usepackage[T1]{fontenc}
\usepackage{cite}
\usepackage{graphicx}
\usepackage{amssymb} 
\usepackage{graphicx}
\usepackage{anyfontsize}
\usepackage{multirow}
\usepackage{booktabs}
\usepackage{tabularx} 
\usepackage{amssymb} 
\usepackage{amssymb}
\usepackage[table]{xcolor}
\usepackage{colortbl}
\usepackage{siunitx}
\usepackage[colorlinks,allcolors=black,pdftex]{hyperref}
\usepackage{array}
\usepackage{pifont}
\usepackage{array} 
\usepackage{multirow}
\usepackage{hyperref}
\newcommand{\xmark}{\ding{55}}%
\definecolor{lighterskyblue}{rgb}{0.8, 0.87, 0.90}
\usepackage{footnote} 
\makesavenoteenv{tabular}
\makesavenoteenv{table}

\newcommand{\myparagraph}[1]{\noindent{\bf #1}}
\newcommand*\samethanks[1][\value{footnote}]{\footnotemark[#1]}

\begin{document}
\title{Parameter Efficient Fine Tuning for Multi-scanner PET to PET Reconstruction}
\titlerunning{PEFT for Multi-scanner PET to PET Reconstruction}
\author{Yumin Kim\thanks{Equal contribution}\and Gayoon Choi\samethanks\and Seong Jae Hwang\thanks{Corresponding author}}
\authorrunning{Y. Kim et al.}
\institute{Department of Artificial Intelligence, Yonsei University, Seoul, Republic of Korea\\
\email{\tt\small \{yumin, gynchoi17, seongjae\}@yonsei.ac.kr}}

\maketitle            
\begin{abstract}
\input{sec/00_abs.tex}
\end{abstract}
\section{Introduction}
\input{sec/01_intro}

\section{Methodology}
\input{sec/03_methods}
\section{Experiments and Results}
\input{sec/04_exp}

\input{sec/05_conc}

\begin{credits}
\subsubsection{\ackname} 
This work was supported in part by the IITP 2020-0-01361 (AI Graduate School Program at Yonsei University), NRF RS-2023-00262002, and NRF RS-2023-00219019 funded by Korean Government (MSIT).

\subsubsection{Disclosure of Interests.} The authors have no competing interests.
\subsubsection{\discintname}

\end{credits}

\bibliography{ref}
\bibliographystyle{splncs04}

\input{sec/Supplementary}
\end{document}

%% file: sec/00_abs.tex
Reducing scan time in Positron Emission Tomography (PET) imaging while maintaining high-quality images is crucial for minimizing patient discomfort and radiation exposure. Due to the limited size of datasets and distribution discrepancy across scanners in medical imaging, fine-tuning in a parameter-efficient and effective manner is on the rise. Motivated by the potential of Parameter-Efficient Fine-Tuning (PEFT), we aim to address these issues by effectively leveraging PEFT to improve limited data and GPU resource issues in multi-scanner setups. In this paper, we introduce \textbf{PETITE}, \textbf{P}arameter-\textbf{E}fficient Fine-\textbf{T}uning for Mult\textbf{I}-scanner PE\textbf{T} to PET R\textbf{E}construction that uses fewer than 1\% of the parameters. To the best of our knowledge, this study is the first to systematically explore the efficacy of diverse PEFT techniques in medical imaging reconstruction tasks via prevalent encoder-decoder-type deep models. This investigation, in particular, brings intriguing insights into PETITE as we show further improvements by treating encoder and decoder separately and mixing different PEFT methods, namely, Mix-PEFT. Using multi-scanner PET datasets comprised of five different scanners, we extensively test the cross-scanner PET scan time reduction performances (i.e., a model pre-trained on one scanner is fine-tuned on a different scanner) of 21 feasible Mix-PEFT combinations to derive optimal PETITE. We show that training with less than 1\% parameters using PETITE performs on par with full fine-tuning (i.e., 100\% parameter).

\keywords{Parameter-Efficient Fine-Tuning \and Vision Transformer  \and Positron Emission Tomography (PET)  \and PET reconstruction}

%% file: sec/01_intro.tex
\begin{figure}[t]
\centering
\includegraphics[width=\textwidth]{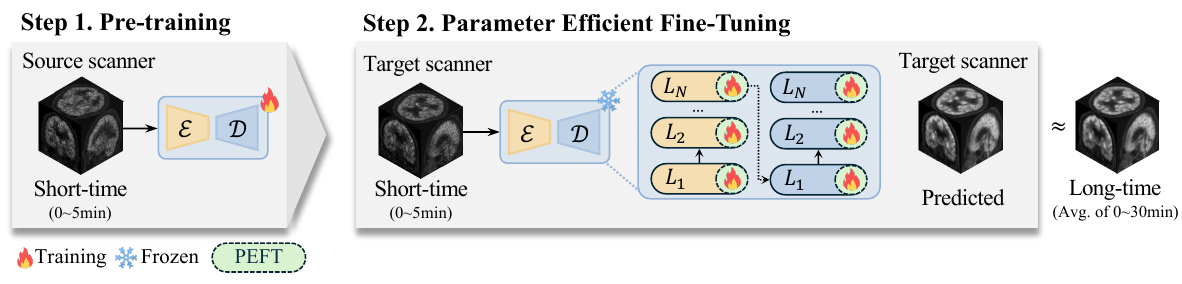} 
\caption{\textbf{Illustration of multi-scanner PET scan time reduction with PEFT}}
\label{fig:fig_str}
\end{figure}

Positron Emission Tomography (PET) is an in vivo nuclear medicine technique using radiotracers for early diagnosis of Alzheimer's and Parkinson's diseases \cite{cummings2019national,becker2002early}.
To address this issue, PET reconstruction emerges as a remarkable technique, enhancing image quality significantly without extending scanning time \cite{conti2011focus, trans-gan, cvtgan}.
It has continuously evolved through various deep learning methods, enabling the reconstruction of high-quality scans from short scans that closely match those acquired from longer scans.

An array of generative models to achieve PET reconstruction have been developed. For instance, Wang \textit{et al.}~\cite{3dcgan} leveraged a Generative Adversarial Network (GAN)~\cite{goodfellow2014generative} to synthesize images that are hard to discriminate with high-dose PET. 
Recently, the rise of Vision Transformer (ViT)~\cite{ViT} has also demonstrated its potential for PET reconstruction. 
Expanding on this integration, Luo \textit{et al.}~\cite{trans-gan} inserts ViTs in CNN Encoder-Decoder to take advantage of both CNN and Transformer and Zeng \textit{et al.}~\cite{cvtgan} adopt convolution in the self-attention mechanism to reduce semantic ambiguity. However, these models encounter efficiency challenges due to the process known as full fine-tuning (Full-FT), which involves updating all layers of a large model, resulting in increased training time and higher GPU resource consumption.

In medical imaging, the diversity of equipment and protocols poses persistent practical challenges, significantly impacting data generalization and model application. This issue manifests similarly in the PET domain due to discrepancies in scanner manufacturers, imaging facilities, or protocol types that complicate application from hospital to hospital.
Such variations limit the generalization of models trained on a particular dataset to others, requiring fine-tuning in hospitals with limited datasets. 
Specifically, among fine-tuning methods, generalization in multi-scanner is also an important issue.

Consequently, due to the scarcity of datasets and inefficient large-scale models, some works progressively adopted \textit{Parameter-Efficient Fine-Tuning} (PEFT).

PEFT approaches significantly decrease storage requirements and computational costs by freezing most parameters of a pre-trained model and selectively fine-tuning a limited set of parameters. 
Despite considerable research on PEFT~\cite{Ptuning, Prefix, kronA}, there are few attempts at medical imaging, with prior studies primarily focusing on classification tasks~\cite{dutt2023parameter, unifiedPEFT}.

Recognizing the gaps in applying PEFT within medical imaging, particularly in reconstruction tasks, we pioneer the use of the PEFT methodology for PET reconstruction from short-time scans, aiming to reduce scan duration and enhance reconstruction quality~\cite{anwar2018medical, AGI}. Specifically in scenarios with relatively limited data availability, we aim to enhance reconstruction performance across diverse scanners using various PEFT methods. Consequently, we introduce \textbf{PETITE}, \textbf{P}arameter-\textbf{E}fficient Fine-\textbf{T}uning for Mult\textbf{I}-scanner PE\textbf{T} to PET R\textbf{E}construction, which represents the most optimal PEFT combination when applying the encoder-decoder approach to each model architecture. PETITE uses fewer than 1\% of the parameters for PET reconstruction using short-time scans aimed at reducing scan time, as detailed in Fig.~\ref{fig:fig_str}, which describes its scheme for single source-target settings. Furthermore, we investigate the optimal experimental settings for reconstruction models through various PEFT approaches.
Specifically, we explore the synergistic effects of applying diverse PEFT methods independently to the distinct encoder and decoder components, the process which is defined as \textbf{Mix-PEFT}. 

\myparagraph{Contributions.} Our main contributions are as follows: \noindent\textbf{(1)} 
We leverage the PEFT methodology in a medical reconstruction task to reduce the scan time of PET images on scanners with different dimensions, voxel spacing, and institutions. 
To the best of our knowledge, this extensive study represents the first application of the PEFT methodology within the field of medical imaging. \textbf{(2)} Upon experimenting with possible Mix-PEFT, we found that using less than 1\% of parameters can achieve performance comparable to Full-FT, carefully considering encoder and decoder architecture.
\textbf{(3)} We provide novel insights into the optimal PEFT settings tailored for the reconstruction model.

%% file: sec/03_methods.tex
\begin{figure}[t]
\centering
\includegraphics[width=\textwidth]{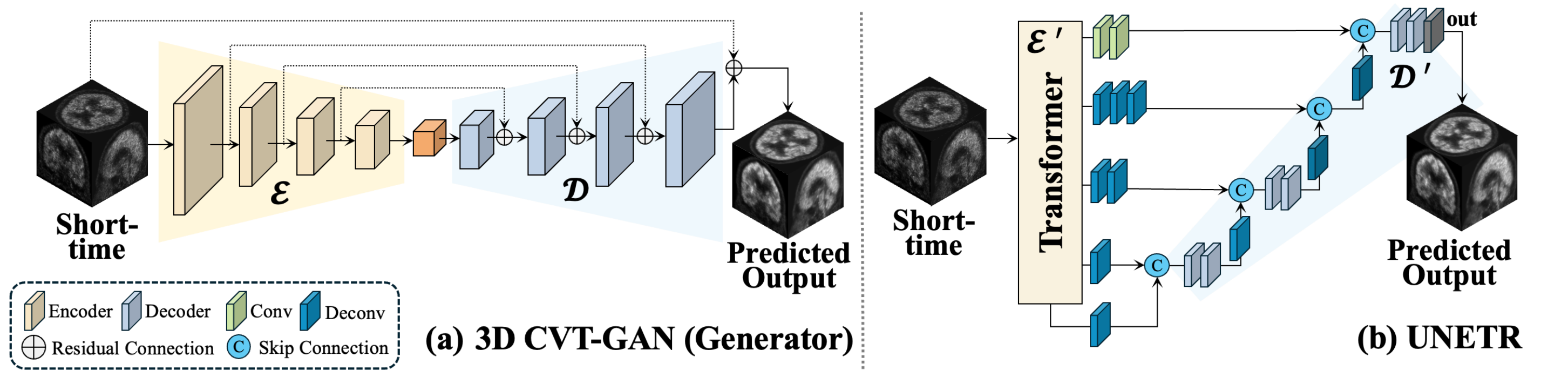} 
\caption{~\textbf{The pipeline of the encoder-decoder structure of each ViT-based model.}
(a) 3D CVT-GAN~\cite{cvtgan} features a generator with a ViT-based encoder and decoder. Only the first three layers of the encoder and the first two layers of the decoder are trained.
(b) UNETR~\cite{unetr} consists of a ViT-based encoder and a CNN-based decoder.}\label{fig:model_arch} 
\end{figure}

We briefly describe relevant PET reconstruction models and outline PEFT approaches (Sec.~\ref{sec:scantime_reduction}). 
Then, we detail Mix-PEFT to consider the encoder and decoder separately (Sec.~\ref{sec:mix_peft}).
\subsubsection{Scan-time Reduction Model.} (a) 3D CVT-GAN~\cite{cvtgan} aims to integrate CNN and ViT technologies for high-quality PET reconstruction effectively. 
This architecture replaces projection in multi-head attention from the linear to the convolutional~\cite{microsoft_cvt}, building encoder-decoder structure and combining the conditional GAN~\cite{cgan}. 
The architecture includes an encoder for feature extraction and a decoder for restoring high-quality PET images, capturing both local spatial features and global contexts from various network layers. 
During the pre-training step, the discriminator is also trained, but it is frozen in the parameter efficient fine-tuning (PEFT) step. 
(b) UNETR~\cite{unetr} comprises a ViT encoder for extracted representations that are merged with the CNN-based decoder through skip connections at multiple resolutions to predicted outputs. 
For detailed structure into the positions and structures of the encoder and decoder in the model used, as shown in Fig.~\ref{fig:model_arch}.

\begin{figure}[t]
\includegraphics[width=\textwidth]{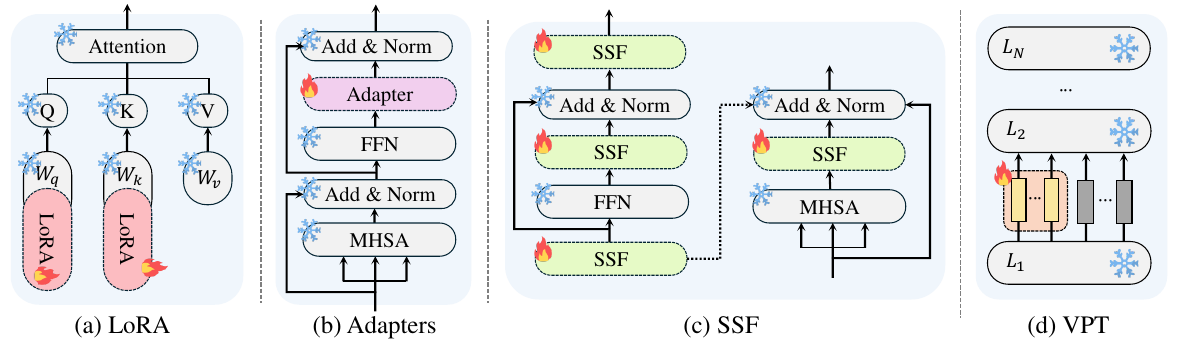}
\caption{\textbf{Illustrations of the modified structures of PEFT methods.}} \label{fig_peft}
\end{figure}
\input{sec/02_PEFT}

\subsection{PETITE: Optimal Effectiveness of Mix-PEFT}
\label{sec:mix_peft}
Mix-PEFT refers to the approach of applying various PEFT methods independently to the encoder and decoder within each model, aiming to achieve and analyze the ideal synergistic effect. 
This method particularly highlights PEFT techniques that consistently provide positive impacts on both the encoder and decoder in two ViT-based models, the 3D CVT-GAN and UNETR models.

\textbf{(\uppercase\expandafter{\romannumeral1}) Vit-based Generator:} PETITE utilizes VPT to the encoder and LoRA to the decoder, respectively, in 3D CVT-GAN. 
By applying task-specific prompt tokens to the encoder via VPT, the 3D CVT-GAN is fine-tuned for specific tasks to improve feature extraction. 
For parameter efficiency, the decoder tunes only the query and key matrices in the attention layer to produce high-quality PET images. Using them together creates a synergistic effect, as both operate within the critical attention component of ViT. 

\textbf{(\uppercase\expandafter{\romannumeral2}) Vit-based encoder and a CNN-based decoder:} PETITE is applying LoRA to the encoder and SSF to the decoder, respectively, in UNETR. 
SSF is best for CNNs. Tuning SSF in ViT with other PEFT methods results in a performance drop, as it hinders ViT's complex attention mechanism. While PEFT was originally designed for ViT layers, it has been adapted to be compatible with CNN-based decoders in our work.
The representations extracted from the encoder are matched to the target dataset's distribution by applying SSF to the CNN-based decoder, utilizing the features of the pre-trained model.

%% file: sec/02_PEFT.tex
\subsection{PEFT for PET Scan-time Reduction}
\label{sec:scantime_reduction}
We describe the PEFT concept and describe the model-specific adjustments and their rationale. We have divided the PEFT category into two variants: Selective and Additive methods. The optimal hyperparameter values for PEFT are given in the Supplementary.

\subsubsection{Selective methods.} This method includes (1) LayerNorm tuning, and (2) BitFit tuning, and leverages the model's pre-training procedure without making major changes. This approach precisely fine-tunes specific layers or a segment of the original pre-trained model, without adding new modules, optimizing performance efficiently.

\textbf{\textit{LayerNorm tuning}} tunes parameters $\theta'\in\theta$ in the normalization layer for adapting intermediate statistics to align with target distributions. 

\textbf{\textit{BitFit}} proposes only updates bias-term of the network. Based on our experimental results, BitFit tuning has been identified as one of the competitive baselines for PEFT. Building on this discovery, we found that combining BitFit tuning with all PEFT methodologies yields better performance than using original PEFT alone. Specifically, integrating BitFit with particularly effective techniques in our study, such as LoRA, SSF, and VPT described below, led to notable performance enhancements, as shown in Table~\ref{ablation}(a). These findings indicate that BitFit and PEFT methods are effectively complementary.

\subsubsection{Additive methods.} The additive method can be categorized into four types~\cite{sdsu}: (1) LoRA, (2) Adapters, (3) SSF, and (4) VPT. This approach involves augmenting the pre-trained model with additional trainable layers, where only the parameters of these new layers are subject to training.

\textbf{\textit{LoRA}} injects learnable rank decomposition matrices into pre-trained weight matrices, under the hypothesis that, during adaptation, weight updates have a low intrinsic rank~\cite{lora}. The low-rank decomposition matrix is scaled with a factor ${\alpha}$ that is constant to a row-rank $r$, and the number of trainable parameters depends on $r$. As shown in Fig.~\ref{fig_peft}(a), we apply LoRA to the query and key matrices, since tuning only these matrices yields better results and a parameter reduction of 0.22\% for the 3D CVT-GAN model and 0.03\% for UNETR, resulting in better performance, with more simplicity than adjusting the query, key, and value matrices together. In UNETR, tuning with a learning rate higher than 1e-2 leads to a fall in local optima and hinders training. Analysis of rank revealed that $r=1,4,8$ are stable in performance, whereas $r=16$ resulted in decreased performance.

\textbf{\textit{Adapters}} introduces lightweight modules, adding fully connected networks after attention, and Feed-forward Network layers in Transformer~\cite{Adapters}. An adapter module typically comprises a linear down-projection, succeeded by a nonlinear activation function, and concluded with a linear up-projection, all integrated with a residual connection. The bottleneck architecture enables parameter reduction through a reduction factor (rf). As seen in Fig.~\ref{fig_peft}(b), we add the Adapter only after the feed-forward layer to further reduce the computational cost according to Pfieffer~\cite{AdapterFusion}. We test adapters sizes in \{1,4,8,16,32\}. Analysis of rank revealed that r = 4,8 are stable in performance.

\textbf{\textit{Scaling \& Shifting Your Features (SSF)}} modulates deep features $x$ from the pre-trained model via linear transformations~\cite{ssf_}. SSF module consists of the scale factor $\gamma$, dot product with $x$, and the shift factor $\beta$, added to $x$. It is injected after the multi-head attention layer, the feed-forward layer, and the normalization layer of the Transformer. As shown in Fig.~\ref{fig_peft}(c), it is injected after every MLP, MHSA, and Layernorm module, scaling and shifting features from them during training, and can be re-parameterized at inference since it is a linear structure. 

\textbf{\textit{Visual-Prompt Tuning (VPT)}} introduces a small set of $p$ continuous vectors in the embedding space of every encoder, tailored to learn task-specific information via attention.~\cite{vpt_}. Since VPT-deep, which inserts prompt tokens into every encoder layer, does not align well with reconstruction tasks, our approach solely utilizes VPT-shallow. Additionally, we observed that inserting prompt tokens from the second encoder, bypassing the first, enhanced performance as depicted in Fig.~\ref{fig_peft}(d). Inserting prompt tokens into the first encoder layer can disrupt representation learning, as it may interfere with focusing on blocks containing task-relevant information due to the variance in locations across pre-trained ViTs~\cite{improving}. In 3D CVT-GAN, we inserted 8 and 32 tokens into the encoder, while in UNETR, we added 50 tokens. The function of VPT is as follows:

\begin{equation}\label{vpt_our}
[\mathbf{Z}_2,\mathbf{E}_2]=L_2([\mathbf{P},\mathbf{E}_1]), \quad
[\mathbf{Z}_i,\mathbf{E}_i]=L_i([\mathbf{Z}_{i-1},\mathbf{E}_{i-1}]),\quad i=3,4,...,N.
\end{equation}

%% file: sec/04_exp.tex
\begin{figure}[t]
\centering
\includegraphics[width=\textwidth]{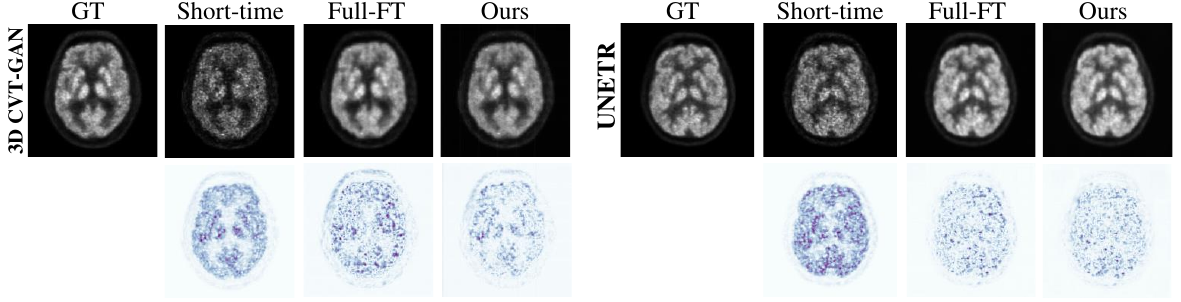}
\caption{\textbf{Scan time reduction examples using 3D CVT-GAN~\cite{cvtgan} and UNETR~\cite{unetr}.} \textit{First row}: PET scans. \textit{Second row}: error maps comparing the reconstructed PET scans to the ground-truth (GT).
}\label{fig: visual}
\end{figure}

\noindent\textbf{Datasets and Image Preprocessing.}
\input{sec/Datasets}

\input{Table/main_result}
\noindent\textbf{Evaluation Metrics.}
\input{sec/preprocessing}

\noindent\textbf{Implementation Details.}
\input{sec/Implementation}

\subsection{Evaluation Results}

\subsubsection{Quantitative Experiments.}
\input{sec/Quantitative_Experiments}
\input{Table/ablation_cvt} 
\subsubsection{Qualitative Experiments.}
\input{sec/Qualitative_Experiments}
\subsubsection{Ablation Study.}
\input{sec/ablations}

%% file: sec/Datasets.tex
In our study, we used the Alzheimer’s Disease Neuroimaging Initiative (ADNI) public dataset for PET imaging across five different scanners and institutions with details on multi-scanner information in Supplementary. $^{18}$F-Fluorodeoxyglucose ($^{18}$F-FDG) was injected at a dose of 185 MBq (5 mCi) for the scans, with each ADNI PET scan consisting of a sequence of six 5-minute frame scans (i.e., 0-5, ..., 25-30 minutes). The short-time scans PET is the first 5-minute scan (0-5) only. The long-time scans PET (GT) is generated by simply averaging these six five-minute frames. Our dataset was split into 30 training and 15 validation samples for pre-training, and 10 training and 15 validation samples for parameter efficient fine-tuning (PEFT).
We processed the images using the MONAI Library, applying a random size crop to $64 \times 64 \times 64$ and normalizing the intensities of all reconstructed PET images to a 0-1 range.

%% file: Table/main_result.tex
\begin{table}[t]
\centering
\caption{\textbf{Quantitative comparison of PSNR, SSIM, and NRMSE  between PETITE and other PEFT methods.} \textit{Best}: \textbf{Bold}; \textit{Second best}: \underline{Underline}.
}
\label{main}
\resizebox{\textwidth}{!}{
\begin{tabular}{>{\raggedright\arraybackslash}l|c|cccc|l|>{\raggedright\arraybackslash}c|cccc}
\hline
\multicolumn{5}{c}{3D CVT-GAN~\cite{cvtgan}} & & \multicolumn{5}{c}{UNETR~\cite{unetr}} \\
\hline
\textbf{Method} & \textbf{\% Param} & \textbf{PSNR}($\uparrow$) & \textbf{SSIM}($\uparrow$) & \textbf{NRMSE}($\downarrow$) & \phantom{a} & \textbf{Method} & \textbf{\% Param} & \textbf{PSNR}($\uparrow$) & \textbf{SSIM}($\uparrow$) & \textbf{NRMSE}($\downarrow$) \\ 
\hline
\multicolumn{1}{l|}{No-FT}          & -       & 30.329 & 0.905 & 0.0327 && \multicolumn{1}{l|}{No-FT}          & -       & 30.034 & 0.787 & 0.0388 \\
\rowcolor[gray]{0.9}Full-FT &  100\% &  \textbf{31.410} &  0.908 &  \textbf{0.0292} && \multicolumn{1}{|>{\raggedright\arraybackslash}l|}{Full-FT} &  100\% & \textbf{31.746} & \textbf{0.884} & \textbf{0.0285} \\
\hline
\multicolumn{1}{l|}{LayerNorm}      & 0.041\%  & 31.304 & \textbf{0.914} & \textbf{0.0292} && \multicolumn{1}{l|}{LayerNorm}      & 0.037\%  & 30.753 & 0.811 & 0.0308 \\
\multicolumn{1}{l|}{BitFit}         & 0.12\%  & 31.291 & 0.910 & \underline{0.0294} && \multicolumn{1}{l|}{BitFit}         & 0.07\%  & 30.969 & 0.810 & 0.0299 \\
\hline
\multicolumn{1}{l|}{LoRA}           & 0.45\%  & 30.982 & 0.909 & 0.0308 && \multicolumn{1}{l|}{LoRA}           & 0.46\%  & 31.010 & 0.825 & 0.0299 \\
\multicolumn{1}{l|}{Adapters}       & 0.71\%  & 31.108 & \underline{0.913} & 0.0304 && \multicolumn{1}{l|}{Adapters}       & 0.44\%  & 30.657 & 0.798 & 0.0311 \\
\multicolumn{1}{l|}{SSF}            & 0.16\%  & 31.018 & 0.809 & 0.0298 && \multicolumn{1}{l|}{SSF}            & 0.11\%  & 30.989 & 0.809 & 0.0298 \\
\multicolumn{1}{l|}{VPT}            & 0.02\%  & 30.398 & 0.892 & 0.0331 && \multicolumn{1}{l|}{VPT}            & 0.038\% & 30.494 & 0.796 & 0.0318 \\
\hline
\rowcolor{lighterskyblue} \multicolumn{1}{l|}{Ours} & 0.32\% & \underline{31.373} & 0.912 & 0.0298 && \multicolumn{1}{|>{\raggedright\arraybackslash}l|}{Ours} & 0.51\% & \underline{31.696} & \underline{0.865} & \underline{0.0292} \\
\hline
\end{tabular}
}
\end{table}

%% file: sec/preprocessing.tex
The peak signal-to-noise ratio (PSNR), structural similarity index measure (SSIM), and normalized root mean squared error (NRMSE) are used as quantitative evaluation metrics. Among them, PSNR indicates estimation accuracy in terms of the logarithmic decibel scale, while SSIM and NRMSE represent the structural similarity and voxel-wise intensity differences between the ground-truth and predicted images, respectively. 

%% file: sec/Implementation.tex
The hyperparameters we tuned include the number of epochs, batch size, learning rate, learning rate decay, learning rate scheduler, the rank value of LoRA~\cite{lora}, the reduction factor of Adapters, and the number of VPT prompt tokens~\cite{vpt_}. 
All models were trained with a batch size of 6, using PyTorch and MONAI for implementation. Both models performed pre-training for 1000 epochs. The PEFT involved training the 3D CVT-GAN for 150 epochs and the UNETR for 200 epochs. The final performance was determined based on the epoch with the highest PSNR value.
See supplementary for detailed hyperparameters for PEFT.

%% file: sec/Quantitative_Experiments.tex
We evaluated our proposed method on two ViT-based models: 3D CVT-GAN~\cite{cvtgan} and UNETR~\cite{unetr}, employing various PEFT methods such as (1) LayerNorm, (2) BitFit, (3) LoRA, (4) Adapters, (5) SSF, (6) VPT, and (7) PETITE (Ours). Specifically, when compared to one of the competitive baselines, Adapters, in the 3D CVT-GAN model (as shown in Table~\ref{main}), our proposed PETITE approach achieved improvements of 0.263 in PSNR and 0.0006 in NRMSE, while utilizing 0.4\% fewer parameters.
Similarly, in the UNETR model, PETITE demonstrated its efficacy as a parameter-efficient approach by improving the metrics PSNR, SSIM, and NRMSE by 1.039, 0.067, and 0.0019, respectively, while maintaining a parameter size comparable to that of Adapters. 

%% file: Table/ablation_cvt.tex
\begin{table}[t]
\centering
\caption{\textbf{Optimal Mix-PEFT (Ours) outperforms each PEFT method with or without BitFit.} \textit{Best}: \textbf{Bold}; \textit{Second best}: \underline{Underline}}
\label{ablation}
\resizebox{\textwidth}{!}{
\begin{tabular}{l|c|c|c|c|c|l|c|c|c|c|c}
\hline
\multicolumn{6}{c|}{\textbf{(a) 3D CVT-GAN~\cite{cvtgan}}} & \multicolumn{6}{c}{\textbf{(b) UNETR~\cite{unetr}}} \\
\hline
\textbf{Method} & \textbf{BitFit} & \textbf{\% Param} & \textbf{PSNR}($\uparrow$) & \textbf{SSIM}($\uparrow$) & \textbf{NRMSE}($\downarrow$) & \textbf{Method} & \textbf{BitFit} & \textbf{\% Param} & \textbf{PSNR}($\uparrow$) & \textbf{SSIM}($\uparrow$) & \textbf{NRMSE}($\downarrow$) \\ 
\hline
No-FT        & -       & -       & 30.329 & 0.905 & 0.0327 & No-FT          & -       & -       & 30.034 & 0.787 & 0.0338 \\
Full-FT         & -       & 100\%   & \textbf{31.410} & 0.908 & \textbf{0.0292} & Full-FT         & -       & 100\%   & \textbf{31.746} & \textbf{0.884} & \textbf{0.0282} \\
\hline
BitFit          & \xmark  & 0.12\%  & 31.291 & 0.910 & \underline{0.0294} & BitFit          & \xmark  & 0.07\%  & 30.969 & 0.810 & 0.0299 \\
VPT             & \xmark  & 0.02\%  & 30.398 & 0.892 & 0.0331 & LoRA (r=4)      & \xmark  & 0.44\%  & 31.010 & 0.825 & 0.0299 \\
LoRA (r=8)      & \xmark  & 0.72\%  & 31.050 & 0.900 & 0.0300 & SSF             & \xmark  & 0.2\%   & 30.897 & 0.808 & 0.0303 \\
\hline
VPT             & \checkmark & 0.14\%  & 31.302 & 0.909 & 0.0300 & LoRA (r=4)      & \checkmark & 0.51\%  & 31.010 & 0.825 & 0.0299 \\
LoRA (r=8)      & \checkmark & 0.61\%  & 31.073 & \underline{0.910} & 0.0307 & SSF             & \checkmark & 0.2\%   & 30.897 & 0.808 & 0.0303 \\
\hline
Ours & \xmark  & 0.20\%  & 31.305 & \underline{0.910} & 0.0307 & Ours & \xmark  & 0.44\%  & 31.193 & 0.819 & 0.0310 \\
\rowcolor{lighterskyblue}
\textbf{Ours} & \cellcolor{lighterskyblue}\checkmark & \cellcolor{lighterskyblue}0.32\%  & \cellcolor{lighterskyblue}\underline{31.373} & \cellcolor{lighterskyblue}\textbf{0.912} & \cellcolor{lighterskyblue}0.0298 & \textbf{Ours} & \cellcolor{lighterskyblue}\checkmark & \cellcolor{lighterskyblue}0.51\%  & \cellcolor{lighterskyblue}\underline{31.696} & \cellcolor{lighterskyblue}\underline{0.865} & \cellcolor{lighterskyblue}\underline{0.0292} \\
\hline
\end{tabular}%
}
\end{table}

%% file: sec/Qualitative_Experiments.tex
In Fig.~\ref{fig: visual} the lighter color of the error map indicates a smaller error. As observed, the quantitative experiments of the 3D CVT-GAN model show that only the PSNR is comparable to the performance of Full-FT Fig.~\ref{fig: visual}, the error is smaller and closer to the GT images compared to Full-FT. 

%% file: sec/ablations.tex
To assess the impact of key components in our proposed PETITE methodology, we carry out ablation studies on two models by considering the following configurations: (1) Baseline, (2) PEFT, (3) PEFT + BitFit, and (4) Ours. 
Commonly, the (1) Baseline encompasses No-FT and Full-FT, while (2) original PEFT methods integrated for the Encoder-Decoder (En-De) structure are proposed in PETITE, and the combination with (3) BitFit is explored to demonstrate the efficacy of tuning the original PEFT alongside BitFit. Although various combinations were possible, (4) Ours represents the best-performing combination identified through ablation studies for each model. 
As shown in Table~\ref{ablation}(a) and Table~\ref{ablation}(b), combining the PEFT method tailored for the En-De structure with BitFit tuning across all layers yields superior performance. Notably, within the UNETR model in Table~\ref{ablation}(b), our approach, despite using 81.9\% fewer parameters than the ViT-based encoder, demonstrates higher performance in terms of PSNR and SSIM by 0.169 and 0.07, respectively, thereby validating the effectiveness of the PETITE methodology.

%% file: sec/05_conc.tex
\section{Conclusions}
Our comprehensive experiments on multi-scanner PET datasets have affirmed the effectiveness of PETITE, demonstrating that the synergetic effects of Mix-PEFT enable achieving results akin to full fine-tuning with less than 1\% of parameters. This efficient approach not only mitigates issues arising from limited datasets and discrepancies in scanner distributions but also addresses the critical need to reduce PET scan times while maintaining image quality. The insights gained from separately addressing encoder and decoder components and integrating various PEFT techniques highlight the potential of PETITE to innovate medical imaging reconstruction tasks. This study lays the groundwork for future research in parameter-efficient methodologies, foreseeing extensive exploration of parameter-efficient fine-tuning methods in the medical imaging field.

%% file: sec/Supplementary.tex
\clearpage

\begin{center}
\normalsize{\textbf{Parameter Efficient Fine Tuning for\\ Multi-scanner PET to PET Reconstruction \\--Supplementary Materials--}}
\end{center} 

\noindent We have described our results in an easily accessible manner on our project page. The link to the project page is as follows: \href{https://micv-yonsei.github.io/petite2024/}{https://micv-yonsei.github.io/petite2024/}.

\begin{table}[h]
\centering 
\caption{\textbf{Details of hyperparameter settings for parameter efficient fine-tuning (PEFT).} We aimed to maintain consistent settings within each PEFT category.}
\label{tab:imple_peft} 
\setlength{\belowrulesep}{0pt}
\setlength{\aboverulesep}{0pt}
\setlength{\tabcolsep}{5pt}

\textbf{(a) 3D CVT-GAN}
\resizebox{\columnwidth}{!}{%
\begin{tabular}{@{}lccccccccc@{}}
\toprule
 & \multicolumn{1}{c}{\multirow{2}{*}{No-FT | Full FT}} & \multicolumn{2}{c}{Selective methods} & \multicolumn{5}{c}{Additive methods} \\             
\cmidrule(lr){3-4} \cmidrule(lr){5-9}
 & & \multicolumn{1}{c}{LayerNorm} & \multicolumn{1}{c}{BitFit} & \multicolumn{1}{c}{LoRA} & \multicolumn{1}{c}{Adapters} & \multicolumn{1}{c}{VPT} & \multicolumn{1}{c}{SSF} & \multicolumn{1}{c}{\textbf{PETITE \textit{(Ours)}}} \\
\midrule
Learning rate & 1e-3 & 1e-3 & 1e-3 & 1e-3 & 1e-2 & 1e-3 & 1e-2 & 1e-3 \\
Weight Decay & 1e-5 & 1e-5 & 1e-5 & 1e-5 & 1e-5 & 1e-5 & 1e-6 & 1e-5 \\
Optimizer & \multicolumn{4}{c}{adam} & \multicolumn{4}{c}{adamw} \\
Learning rate schedule & \multicolumn{8}{c}{CosineAnneal} \\
Total epochs & \multicolumn{8}{c}{150} \\
\bottomrule
\end{tabular}%
}

\textbf{(b) UNETR}
\resizebox{\columnwidth}{!}{%
\begin{tabular}{@{}lccccccccc@{}}
\toprule
 & \multicolumn{1}{c}{\multirow{2}{*}{No-FT | Full FT}} & \multicolumn{2}{c}{Selective methods} & \multicolumn{5}{c}{Additive methods} \\ 
\cmidrule(lr){3-4} \cmidrule(lr){5-9}
 & & \multicolumn{1}{c}{LayerNorm} & \multicolumn{1}{c}{BitFit} & \multicolumn{1}{c}{LoRA} & \multicolumn{1}{c}{Adapters} & \multicolumn{1}{c}{VPT} & \multicolumn{1}{c}{SSF} & \multicolumn{1}{c}{\textbf{PETITE \textit{(Ours)}}} \\
\midrule
Learning rate & 1e-3 & 1e-3 & 1e-3 & 1e-3 & 1e-2 & 1e-2 & 1e-2 & 1e-3 \\
Weight Decay & 1e-5 & 1e-5 & 1e-5 & 1e-4 & 1e-4 & 1e-5 & 1e-5 & 1e-5 \\
Optimizer & \multicolumn{8}{c}{adamw} \\
Learning rate schedule & \multicolumn{6}{c}{WarmupCosine} & \multicolumn{2}{c}{CosineAnneal} \\
Total epochs & \multicolumn{8}{c}{200} \\
\bottomrule
\end{tabular}%
}
\end{table}

\begin{table}
\centering
\caption{\textbf{Scanner specifications.}}\label{dataset_distribution}
\resizebox{\textwidth}{!}{%
\begin{tabular}{c|@{\hskip\tabcolsep}cccc}
\toprule
Scanner & Resolution & Voxel spacing & Manufacturer & Institution \\
\hline
1 & (192, 192, 136) & (1.21875, 1.21875, 1.21875) & Siemens & Univ of California \\
2 & (192, 192, 128) & (1.21875, 1.21875, 1.21875) & Siemens & Univ of California \\
3 & (224, 224, 81) & (1.01821, 1.01821, 2.02699) & Siemens & Univ of California \\
4 & (128, 128, 90) & (2, 2, 2) & Philips Healthcare & OHSU \\
5 & (128, 128, 63) & (2.05941, 2.05941, 2.425) & Siemens & UCSD \\
\bottomrule
\end{tabular}%
}
\end{table}

\begin{table}[]
\centering
\caption{\textbf{Computation costs of PEFT methods.}}
\begin{tabular}{lcccccccc}
\toprule
\multicolumn{1}{c|}{Method}                    & Full-FT & LN           & BitFit & LoRA  & Adapters & SSF   & VPT            & \textbf{PETITE \textit{(Ours)}} \\ \hline
\multicolumn{9}{c}{\textbf{(a) 3D CVT-GAN}}                                                                                                           \\ \hline
\multicolumn{1}{l|}{Train time (Sec)}          & 450     & 360          & 355    & 340   & 360      & 420   & 398            & \textbf{307}           \\
\multicolumn{1}{l|}{GPU memory (MB)}           & 8,430   & 7,899        & 6,092  & 6,416 & 6,075    & 8,225 & 8,066          & \textbf{4,725}         \\ \hline
\multicolumn{9}{c}{\textbf{(b) UNETR}}                                                                                                                \\ \hline
\multicolumn{1}{l|}{Train time (Sec)} & 580     & \textbf{330} & 417    & 480   & 420      & 432   & 424            & 409                    \\
\multicolumn{1}{l|}{GPU memory (MB)}           & 8,430   & 6,525        & 7,805  & 7,812 & 8,026    & 4,918 & \textbf{4,875} & 5,940                  \\ \bottomrule
\end{tabular}
\end{table}

\begin{table}[t]
\setlength{\belowrulesep}{0pt}
\setlength{\aboverulesep}{0pt}
\setlength{\tabcolsep}{5pt}
\centering
\caption{\textbf{Feasible Mix-PEFT combinations with BitFit for 3D CVT-GAN.}}
\resizebox{\columnwidth}{!}{%
\begin{tabular}{@{}c|c|c|ccc@{}}
\toprule
\textbf{Vit-based Encoder}         & \textbf{Vit-based Decoder}         & \textbf{\%Param} & \textbf{PSNR}($\uparrow$)   & \textbf{SSIM}($\uparrow$)   & \textbf{NRMSE}($\downarrow$) \\ \midrule
LoRA ($r=8$)      & Adapters ($rf=8$) & $0.69\%$    & $30.870$ & $0.908$ & $0.0394$ \\
                & SSF             & $0.58\%$    & $30.556$ & $0.862$ & $0.0403$ \\
                & VPT ($32/8$)      & $0.28\%$    & $30.970$ & $0.908$ & $0.0380$ \\ \midrule
Adapters ($rf=8$) & LoRA ($r=8$)      & $0.69\%$    & $29.956$ & $0.823$ & $0.0343$ \\
                & SSF             & $0.78\%$    & $30.821$ & $0.860$ & $0.0312$ \\
                & VPT ($32/8$)      & $0.87\%$    & $31.203$ & $0.910$ & $0.0302$ \\ \midrule
\cellcolor{lighterskyblue}\textbf{VPT ($0/8/32$)}            & \cellcolor{lighterskyblue}\textbf{LoRA ($r=8$)}           & \cellcolor{lighterskyblue}\textbf{$0.32\%$}    & \cellcolor{lighterskyblue}\textbf{$31.373$} & \cellcolor{lighterskyblue}\textbf{$0.912$} & \cellcolor{lighterskyblue}\textbf{$0.0298$} 
     \\
                & Adapters ($rf=8$) & $0.28\%$    & $31.331$ & $0.910$ & $0.0300$ \\
                & SSF             & $0.46\%$    & $30.059$ & $0.826$ & $0.0433$ \\ \midrule
SSF             & LoRA ($r=8$)      & $0.35\%$    & $30.522$ & $0.812$ & $0.0454$ \\
                & Adapters ($rf=8$) & $0.29\%$    & $30.728$ & $0.882$ & $0.0332$ \\
                & VPT ($32/8$)      & $0.38\%$    & $31.314$ & $0.904$ & $0.0314$ \\ \bottomrule

\end{tabular}%
}
\end{table}

\begin{table}[t]
\setlength{\belowrulesep}{0pt}
\setlength{\aboverulesep}{0pt}
\setlength{\tabcolsep}{5pt}
\centering
\caption{\textbf{Feasible Mix-PEFT combinations with BitFit for UNETR.}}
\label{tab:my-table}
\resizebox{\columnwidth}{!}{%
\begin{tabular}{@{}c|c|c|ccc@{}}
\toprule
\textbf{Vit-based Encoder}         & \textbf{CNN-based Decoder}         & \textbf{\%Param} & \textbf{PSNR}($\uparrow$)   & \textbf{SSIM}($\uparrow$)   & \textbf{NRMSE}($\downarrow$) \\ \midrule
\hline
\cellcolor{lighterskyblue}\textbf{LoRA ($r=4$)}      & Adapters ($rf=4$) & $0.51\%$    & $31.545$ & $0.860$ & $0.0311$ \\
                & \cellcolor{lighterskyblue}\textbf{SSF}      & \cellcolor{lighterskyblue}{\textbf{$0.51\%$}}    & \cellcolor{lighterskyblue}{\textbf{$31.696$}} & \cellcolor{lighterskyblue}{\textbf{$0.865$}} & \cellcolor{lighterskyblue}{\textbf{$0.0292$}} \\ \midrule
Adapters ($rf=16$) & LoRA ($r=4$)    & $0.64\%$    & $30.921$ & $0.839$ & $0.0340$ \\
                & SSF      & $0.95\%$    & $31.670$ & $0.864$ & $0.0310$ \\ \midrule
VPT ($50$)            & LoRA ($r=4$)     & $0.53\%$    & $31.193$ & $0.819$ & $0.0290$ \\
                & Adapters ($rf=4$) & $0.42\%$    & $31.06$  & $0.832$ & $0.0335$ \\
                & SSF      & $0.11\%$    & $31.137$ & $0.841$ & $0.0339$ \\ \midrule
SSF             & Adapters ($rf=4$) & $0.28\%$    & $31.207$ & $0.840$ & $0.0342$ \\
                & LoRA ($r=4$)     & $0.52\%$    & $31.123$ & $0.841$ & $0.0294$ \\ \bottomrule
\end{tabular}%
}
\end{table}
